\documentclass[aps,prd,reprint,preprintnumbers,amsmath,amssymb,floatfix,superscriptaddress]{revtex4-1}

\usepackage{graphicx}
\usepackage{epsfig}
\usepackage{dcolumn}
\usepackage{bbm}
\usepackage[colorlinks,linkcolor=red,anchorcolor=green,citecolor=blue,CJKbookmarks=True]{hyperref}
\usepackage{braket}

\usepackage{multirow} 

\newcommand{\red}[1]{\textcolor{red}{#1}}

\begin{document}
\title{Form factor for Dalitz decays from $J/\psi$ to light pseudoscalars}

\author{\small Chunjiang Shi}
\email{shichunjiang@ihep.ac.cn}
\affiliation{Institute of High Energy Physics, Chinese Academy of Sciences, Beijing 100049, People's Republic of China}
\affiliation{School of Physical Sciences, University of Chinese Academy of Sciences, Beijing 100049, People's Republic of China}

\author{\small Ying Chen}
\email{cheny@ihep.ac.cn}
\affiliation{Institute of High Energy Physics, Chinese Academy of Sciences, Beijing 100049, People's Republic of China}
\affiliation{School of Physical Sciences, University of Chinese Academy of Sciences, Beijing 100049, People's Republic of China}

\author{\small Xiangyu Jiang}
\affiliation{Institute of High Energy Physics, Chinese Academy of Sciences, Beijing 100049, People's Republic of China}
\affiliation{CAS Key Laboratory of Theoretical Physics, Institute of Theoretical Physics, Chinese Academy of Sciences, Beijing 100190, People's Republic of China}

\author{\small Ming Gong}
\affiliation{Institute of High Energy Physics, Chinese Academy of Sciences, Beijing 100049, People's Republic of China}
\affiliation{School of Physical Sciences, University of Chinese Academy of Sciences, Beijing 100049, People's Republic of China}

\author{\small Zhaofeng Liu}
\affiliation{Institute of High Energy Physics, Chinese Academy of Sciences, Beijing 100049, People's Republic of China}
\affiliation{Center for High Energy Physics, Peking University, Beijing 100871, People's Republic of China}

\author{\small Wei Sun}
\affiliation{Institute of High Energy Physics, Chinese Academy of Sciences, Beijing 100049, People's Republic of China}

\def\modified#1{\red{#1}}

\begin{abstract}
We calculate the form factor $M(q^2)$ for the Dalitz decay $J/\psi\to \gamma^*(q^2)\eta_{(N_f=1)}$ with $\eta_{(N_f)}$ being the SU($N_f$) flavor singlet pseudoscalar meson. The difference among the partial widths $\Gamma(J/\psi\to \gamma \eta_{(N_f)})$ at different $N_f$ can be attributed in part to the $\mathbf{U}_A(1)$ anomaly that induces a $N_f$ scaling. $M(q^2)$'s in $N_f=1,2$ are both well described by the single pole model $M(q^2)=M(0)/(1-q^2/\Lambda^2)$. Combined with the known experimental results of the Dalitz decays $J/\psi\to Pe^+e^-$, the pseudoscalar mass $m_P$ dependence of the pole parameter
$\Lambda$ is approximated by $\Lambda(m_P^2)=\Lambda_1(1-m_P^2/\Lambda_2^2)$ with $\Lambda_1=2.64(4)~\mathrm{GeV}$ and $\Lambda_2=2.97(33)~\mathrm{GeV}$. These results provide inputs for future theoretical and experimental studies on the Dalitz decays $J/\psi\to Pe^+e^-$.

\end{abstract}
\maketitle
\section{Introduction}
The electromagnetic (EM) Dalitz decay of a hadron $A$, namely, $A\to B\gamma^*\to B l^+l^-$, refers to the decay process that $A$ decays into $B$ by emitting a time-like photon which then converts to a lepton pair $l^+l^-$. The differential partial decay width with respect to the invariant mass $q^2\equiv m^2_{l^+l^-}$ of the lepton pair can be expressed by $d\Gamma(q^2)/dq^2=(d\Gamma(q^2)/dq^2)_\mathrm{point-like} |f_{AB}(q^2)|^2$~\cite{Landsberg:1985gaz}, where $(d\Gamma(q^2)/dq^2)_\mathrm{point-like}$ can be calculated exactly in QED for point-like particle $A$ and $B$ and the $f_{AB}(q^2)$ is called the transition form factor (TFF) of the transition $A\to B$, and is an important probe to the EM structure of the $AB\gamma$ vertex and also the internal structure of the hadron $A$ (and $B$ if it is also structured). Experimentally, the TFF $f_{AB}(q^2)$ can be derived by taking the ratio $[d\Gamma(q^2)/dq^2]/\Gamma(A\to B\gamma)\propto |f_{AB}(q^2)|^2/|f_{AB}(0)|^2\equiv|F_{AB}(q^2)|^2$ with the normalization $F_{AB}(0)=1$, where many systematic uncertainties cancel. The Dalitz decays of light hadrons, such as $\phi\to \pi^0 e^+e^-$~\cite{Achasov:2002hs,KLOE-2:2016pnx}, $\phi\to \eta e^+e^-$~\cite{KLOE-2:2014hfk}, $\omega\to \pi^0 e^+e^-$~\cite{CMD-2:2005dew}, $\omega\to \pi^0 \mu^+\mu^-$~\cite{Dzhelyadin:1979yw}, have been widely studied in experiments. It should be noted that the experimental studies on Dalitz decays usually require large statistics, since they are rare decays for a hadron.     

The BESIII Collaboration (BESIII) has accumulated more than $10^{10}$ $J/\psi$ events~\cite{BESIII:2021cxx}, based on which the Dalitz decays of $J/\psi$ to light hadrons can be researched. On the other hand, the light pseudoscalars (P), such as $\eta,\eta',\eta(1405/1475)$ and $X(1835)$ etc., are observed to have large production rates on the $J/\psi$ radiative decays~\cite{ParticleDataGroup:2022pth}. So it is expected the Daltiz decays $J/\psi \to Pl^+l^-$ can be investigated to a high precision. Actually, BESIII has performed the experimental studies on the processes $J/\psi\to e^+e^-(\eta,\eta',\pi^0) $~\cite{BESIII:2014dax}, $J/\psi\to e^+e^-\eta$~\cite{BESIII:2018qzg}, $J/\psi\to e^+e^-\eta'$~\cite{BESIII:2018iig,BESIII:2018aao}, $J/\psi\to e^+e^- \eta(1405)$~\cite{BESIII:2023jyg} and $J/\psi\to e^+e^-(X(1835),X(2120),X(2370)$~\cite{BESIII:2021xoh}. With the large $\psi(3686)$ data ensemble, BESIII also studies the process $\psi(3686)\to e^+e^- \eta_c$. The TTFs $F_{\psi p}(q^2)$ are extracted for the processes $J/\psi\to \eta,\eta',\eta(1405),X(1835)$ and $\psi(3686)\to \eta_c$, and the $q^2$-dependence can be described by the single-pole model 
\begin{equation}\label{eq:pole}
    F_{\psi P}=\frac{1}{1-q^2/\Lambda^2},
\end{equation}
based on the vector meson dominance (VMD)~\cite{Landsberg:1985gaz,Fu:2011yy,Gu:2019qwo}, and the pole parameter $\Lambda$ varies in the range from 1.7 to 3.8 GeV. 

Intuitively, the Dalitz decay and the radiative decay of charmonium into light pseudoscalars happen through the annihilation of the charm quark and antiquark. According to the OZI rule, the dominant contribution comes from the initial state radiation of the virtual and real photons from the charm (anti)quark. In this sense, the TFF of $J/\psi$ to pseudoscalars should reflect the electromagnetic properties of $J/\psi$. Therefore, For a same initial vector charmonium, Dalitz decays are insensitive to the properties of the final state light hadrons. A theoretical derivation of $F_{\psi P}(q^2)$ from QCD is desirable but is still challenging since $F_{\psi P}(q^2)$ is obviously in the non-perturbative regime of QCD. The phenomenological studies of $F_{\psi \eta^{(')}}(q^2)$ can be found in Ref.~\cite{He:2020jvj} where the analysis is carried out in the full kinematic region based on QCD models and in Ref.~\cite{Chen:2014yta} where the $J/\psi\to\gamma^* P$ is discussed within the framework of the effective Lagrangian approach and the $\eta_c-\eta-\eta'$ mixing is considered. 

Lattice QCD may take the mission to give reliable predictions of TFF of $J/\psi$ to light hadrons. A recent $N_f=2$ lattice QCD calculation confirms the large production rate of the flavor singlet pseudoscalar meson $\eta_{(2)}$ in the $J/\psi$ radiative decay~\cite{Jiang:2022gnd}. In that work, the EM form factor is obtained at quite a few values of time-like $q^2$, from which the on-shell form factor at $q^2=0$ is obtained through a polynomial interpolation. By assuming the $\mathrm{U}_A(1)$ anomaly dominance and using the $\eta-\eta'$ mixing angle, this on-shell form factor results in the branching fractions of $J/\psi\to\gamma\eta$ and $J/\psi \to \gamma \eta'$ that are close to the experimental values. Actually, the $q^2$ dependence of this decay form factor is described better by the single-pole model in Eq.~(\ref{eq:pole}) (see below). 

Recently, we generated a large gauge ensemble with $N_f=1$ strange sea quarks. The $N_f=1$ QCD is a well defined theory and a simplified version of QCD. It has no chiral symmetry breaking but the $\mathrm{U}_A(1)$ anomaly that has a close relation with the unique light pseudoscalar meson $\eta_{(1)}$. So we will revisit the production rate of $\eta_{(1)}$ in the $J/\psi$ radiative decays. We will test the $\mathrm{U}_A(1)$ anomaly dominance in this process by looking at the $N_f$ dependence of the partial decay width, since the $\mathrm{U}_A(1)$ anomaly is proportional to $N_f$. In the meantime, we will explore the $q^2$-dependence of the related TFF and its sensitivity to the light pseudoscalar mass since our sea quark is much heavier than that in Ref.~\cite{Jiang:2022gnd}. The related calculations involve necessarily the annihilation effect of strange quarks which are dealt with using the distillation method~\cite{Peardon:2009gh}.  

This work is organized as follows. The numerical procedures and results are presented in Sec.~\ref{sec:numerical}. Section~\ref{sec:discussion} is devoted to the discussions and the physical implications of our results. Sect.~\ref{sec:summary} is the summary of this work. 

\section{Numerical details}\label{sec:numerical}
\subsection{$N_f=1$ Gauge Ensemble}\label{sec:ensemble}
We generate gauge configurations with $N_f=1$ dynamical strange quarks on an $L^3\times T=16^3\times 128$ anisotropic lattice. We use the tadpole-improved Symanzik's gauge action for anisotropic lattices~\cite{Morningstar:1997ff,Chen:2005mg} and the tadpole-improved anisotropic clover fermion action~\cite{Zhang:2001in,Su:2004sc}. The RHMC algorithm implemented in Chroma software~\cite{Edwards:2004sx} is used to generate the $N_f=1$ gauge configurations. The parameters in the action are tuned to give the anisotropy $\xi=a_s/a_t\approx 5$, where $a_t$ and $a_s$ are the temporal and spatial lattice spacings, respectively. The scale setting takes the following procedure. Experimentally, there is an interesting relation between pseudoscalar meson masses $m_{PS}$ and the vector meson masses $m_V$ of the quark configuration $q_l\bar{q}$,
\begin{equation}\label{eq:dmsq}
    \Delta m^2 \equiv m_V^2 -m_{PS}^2\approx 0.56-0.58~~\mathrm{GeV}^2
\end{equation}
where $q_l$ stands for the $u,d,s$ quarks and $q$ stands for $u,d,s,c$ quarks. The masses of these vector and pseudoscalar mesons from PDG ~\cite{ParticleDataGroup:2022pth} are listed in Table~\ref{tab:v2-ps2} along with their mass squared differences. So we assume the relation of Eq.~(\ref{eq:dmsq}) is somewhat general for light mesons and use it to set the scale parameter $a_t$. 
We make the least squares fitting to the mass squared differences over the $n\bar{n}$, $n\bar{s}$, $n\bar{c}$ and $s\bar{c}$ systems where $n$ refers to the $u,d$ quarks, and get the mean value $\overline{\Delta m^2}=0.568(8)$ $\mathrm{GeV}^2$, which serves as an input to give the lattice scale parameter $a_t^{-1}=6.66(5)$ GeV. Since the HPQCD collaboration determines the 
$s\bar{s}$ pseudoscalar meson mass to be $m_{\eta_s}=0.686(4)$ GeV from the connected quark diagram~\cite{Davies:2009tsa}, we use the ratio 
$m_\phi/m_{\eta_s}=1.487(9)$ to set the bare mass parameters of strange quarks. Even though $\eta_s$ is not a physical state, the mass squared difference $m_\phi^2-m_{\eta_s}^2\approx 0.570 ~\mathrm{GeV}^2$ also satisfies the empirical relation of Eq.~(\ref{eq:dmsq}). Finally, we obtain  $m_{\eta_s}=693.1(3)(6.0)$ MeV, $m_\phi=1027.2(5)(7.7)$ MeV and $m_\phi^2-m_{\eta_s}^2=0.570~\mathrm{GeV}^2$ on our gauge ensemble. This serves as a self-consistent check of our lattice setup. The details of the gauge ensemble are given in Table~\ref{tab:config}. For the valence charm quark, we use the same fermion action as the strange sea quarks and the charm quark mass parameters are tuned to give $(m_{\eta_c}+3m_{J/\psi})/4=3069$ MeV.

The quark propagators are calculated in the framework of the distillation method~\cite{Peardon:2009gh}. Let $\{V_i,i=1,2,\ldots, 70\}$ be the set of the $N_V=70$ eigenvectors (with smallest eigenvalues) of the gauge covariant Laplacian operator on the lattice.
We use these eigenvectors to calculate the perambulators of strange and charm quarks, which are encoded with the all-to-all quark propagators and facilitate the treatment of quark disconnected diagrams. In the meantime, these eigenvectors provide a smearing scheme for quark fields, namely, $\psi^{(s)}=V^\dagger V \psi$, where $\psi^{(s)}$ is the smeared quark field of $\psi$, $V$ is a matrix with each column being an eigenvector. All the meson interpolation operators in this work are built from the smeared charm and strange quark fields.

\begin{table}[t]
    \centering\caption{\label{tab:v2-ps2} Experimental values of the masses of the pseudoscalar (P) and vector mesons (V) of quark configurations $n\bar{q}$, $n\bar{s}$, $n\bar{c}$, $s\bar{c}$, $n\bar{b}$ and $s\bar{b}$~\cite{ParticleDataGroup:2022pth}. Here $n$ refers to the light $u,d$ quarks. The right most column lists the $m_V^2-m_{PS}^2~(\mathrm{GeV}^2)$. In the row of $s\bar{s}$ states, the mass of the $s\bar{s}$ pseudoscalar $\eta_s$ is determined by the HPQCD collaboration from lattice QCD calculations~\cite{Davies:2009tsa}.}
    \begin{ruledtabular}
    \begin{tabular}{cccc}
            $q_l\bar{q}$ & $m_V$ (GeV) & $m_{PS}$ (GeV)              & $m_V^2-m_{PS}^2~(\mathrm{GeV}^2)$ \\
            \hline
            $n\bar{n}$   & 0.775       & 0.140                       & 0.581                             \\
            $n\bar{s}$   & 0.896       & 0.494                       & 0.559                             \\
            $s\bar{s}$   & 1.020       & 0.686~\cite{Davies:2009tsa} & 0.570                             \\
            $n\bar{c}$   & 2.010       & 1.870                       & 0.543                             \\
            $s\bar{c}$   & 2.112       & 1.968                       & 0.588                             \\
            $n\bar{b}$   & 5.325       & 5.279                       & 0.481                             \\
            $s\bar{b}$   & 5.415       & 5.367                       & 0.523                             \\                  
        \end{tabular}
    \end{ruledtabular}
\end{table}

\begin{table}[t]
    \renewcommand\arraystretch{1.5}
    \caption{Parameters of the gauge ensemble.}
    \label{tab:config}
    \begin{ruledtabular}
        \begin{tabular}{lcccccc}
            $L^3 \times T$    & $\beta$ & $a_t^{-1}$(GeV) & $\xi$      & $m_{\eta_s}$(MeV)& $m_\phi$ (MeV) & $N_\mathrm{cfg}$ \\\hline
            $16^3 \times 128$ & 2.0     & $6.66(5)$     & $\sim 5.0$ & $693(5)$  & $1027(8)$ & $1547$           \\
        \end{tabular}
    \end{ruledtabular}
\end{table}

\subsection{Pseudoscalar meson $\eta_{(1)}$}

\begin{figure*}[t]
    \centering
    \includegraphics[width=0.3\linewidth]{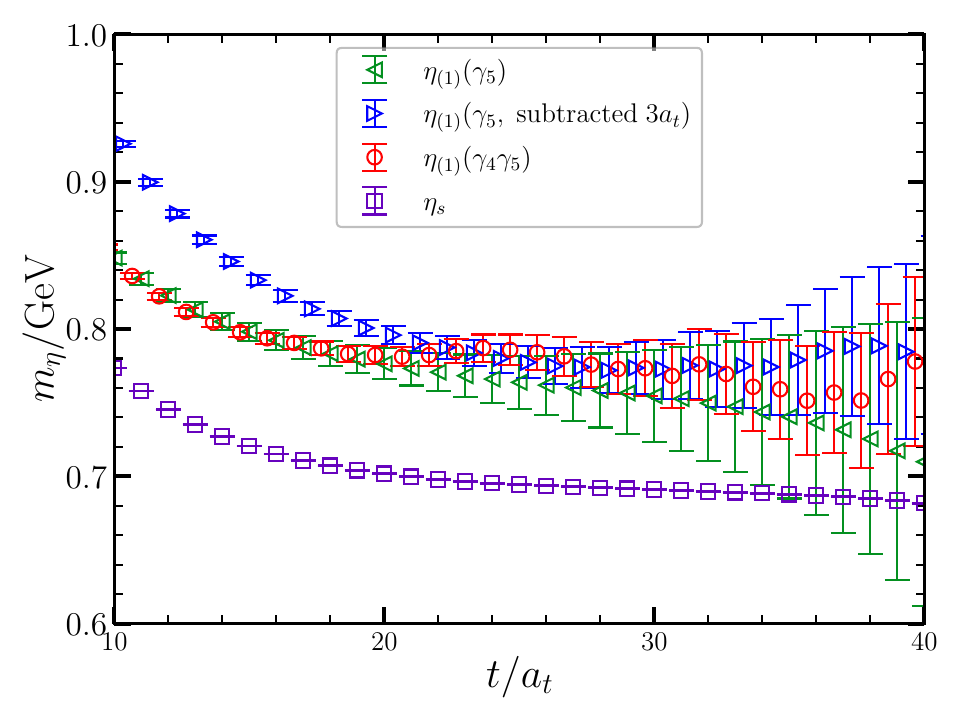}
    \includegraphics[width=0.3\linewidth]{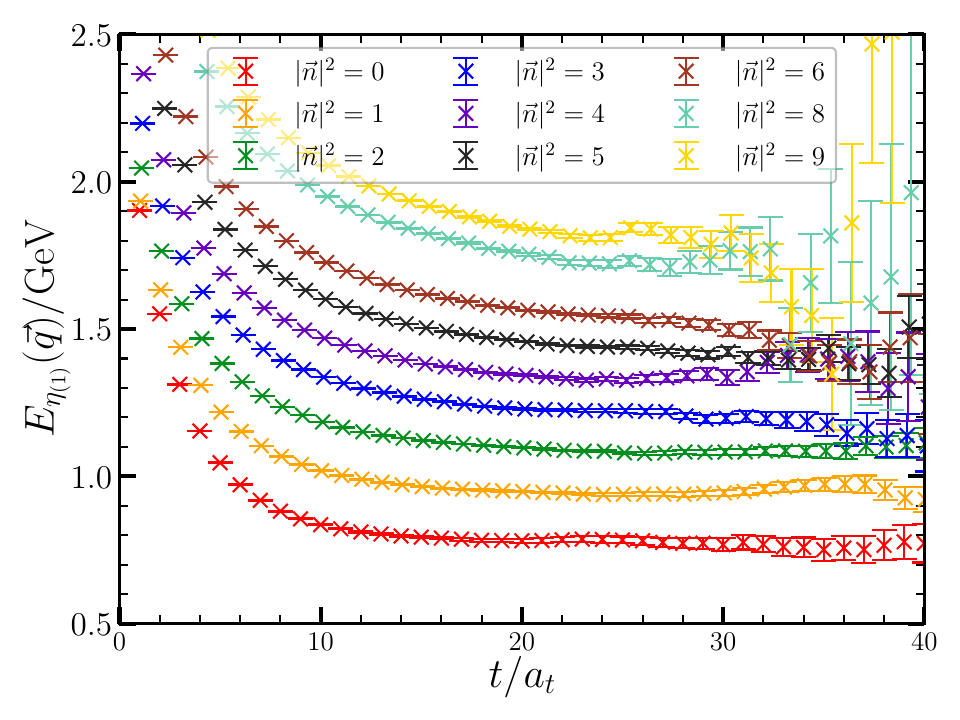}
    \includegraphics[width=0.3\linewidth]{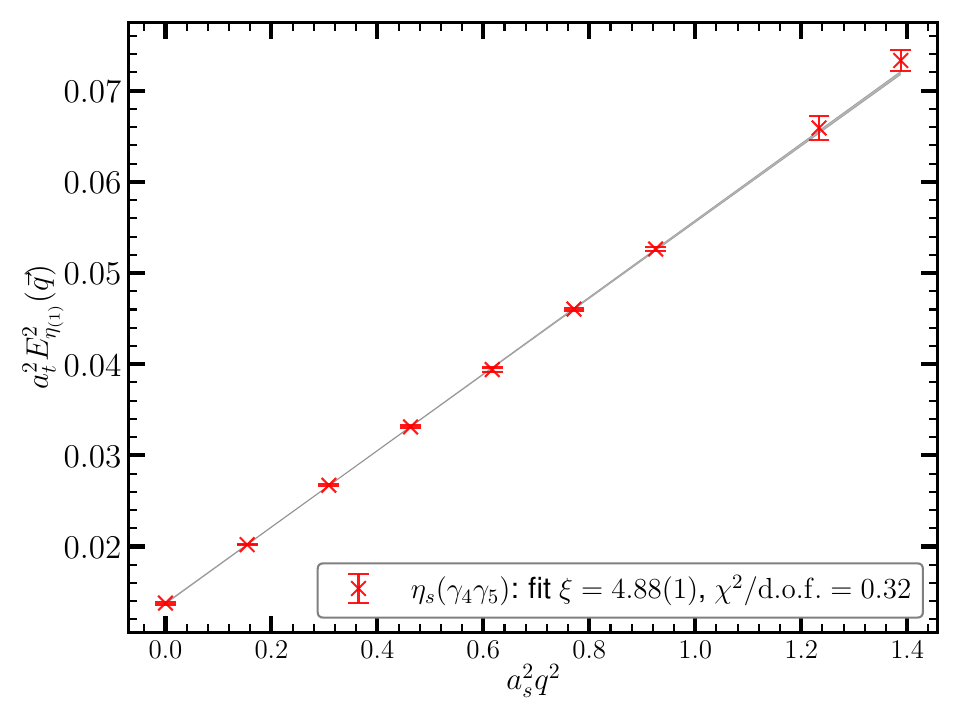}
    \caption{Lattice results of the energies of $\eta_{(1)}$. Left panel: the effective mass of two-point correlation functions of $\eta_{(1)}$ and $\eta_{s}$, where $\eta_{(1)}$ using the operators $\mathcal{O}_{\gamma_5}$ and $\mathcal{O}_{\gamma_4\gamma_5}$. The green points are from the original two-point function of $\mathcal{O}_{\gamma_5}$, the blue points are from the subtracted two-point function in Eq.~(\ref{eq:subtracted}) with $t_0=3 a_t$, and the red points are from the two-point function of $\mathcal{O}_{\gamma_4\gamma_5}$. The plateau regions of red and blue points merge together in large $t/a_t$ range. 
    The purple points stand for unphysical $\eta_s$, which only include connected diagram.
    Middle panel: the effective energies $E_{\eta_{(1)}}^{(\vec{q})}(t)$ with the momentum mode $\vec{n}$ of $\vec{q}$ up to $|\vec{n}|^2=9$, where the data points are from the correlation functions of $\mathcal{O}_{\gamma_4\gamma_5}(\vec{q})$. The fitted $E_{\eta_{(1)}}^{(\vec{q})}$ values are given in Table~\ref{tab:fit-values-energy-form-factor}. Right panel: the dispersion relation of $\eta_{(1)}$. The grey band illustrate the dispersion relation in Eq.~(\ref{eq:dispersion}) with the fitted $\xi=4.88(1)$ and the $\chi^2$ per degree of freedom $\chi^2/\mathrm{d.o.f}=0.32$.}
    \label{fig:effmass}
\end{figure*}

We use two interpolation operators for $\eta_{(1)}$, namely $\mathcal{O}_{\gamma_5}=\bar{s}^{(s)}\gamma_5 s^{(s)}$ and $\mathcal{O}_{\gamma_4\gamma_5}=\bar{s}^{(s)}\gamma_4\gamma_5 s^{(s)}$ to calculate the correlation functions $C_{\gamma_5\gamma_5}(t)$ and $C_{(\gamma_4\gamma_5)(\gamma_4\gamma_5)}(t)$. $C_{\gamma_5\gamma_5}(t)$ has a finite volume artifact that it approaches to a nonzero constant when $t$ is large, as shown in Fig.~\ref{fig:effmass}. This artifact comes from the topology of QCD vacuum and can be approximately expressed as $a^5(\chi_\mathrm{top} + Q^2/V)/T$ where $a$ is the lattice spacing (in the isotropic case), $\chi_\mathrm{top}$ is the topological susceptibility, $Q$ is the topological charge, $V$ is the spatial volume and $T$ is the temporal extension of the lattice~\cite{Aoki:2007ka,Bali:2014pva,Dimopoulos:2018xkm}. In contrast, $C_{(\gamma_4\gamma_5)(\gamma_4\gamma_5)}(t)$ damps to zero for large $t$, which is the normal large $t$ behavior. The constant term of $C_{\gamma_5\gamma_5}(t)$ can be subtracted by taking the difference 
\begin{equation}\label{eq:subtracted}
    C_{\gamma_5\gamma_5}'(t)=C_{\gamma_5\gamma_5}(t)-C_{\gamma_5\gamma_5}(t+t_0), 
\end{equation}
and we take $t_0=3 a_t$ in practice. The effective mass functions
    $m_\mathrm{eff}(t)=\ln \frac{C_{\Gamma\Gamma}^{(')}(t)}{C_{\Gamma\Gamma}^{(')}(t+1)}$
of the two correlations are shown in Fig.~\ref{fig:effmass}, where one can see that decent mass plateaus show up when $t/a_t>15$ and agree with each other. The effective masses of the connected parts of the two correlation functions are also shown for comparison. Their plateaus correspond to the 
mass $m_{\eta_s}$ of $\eta_s$. The data analysis gives the results
\begin{equation}
    {m}_{\eta_s}=693.1(3)~\text{MeV}, ~~~m_{\eta_{(1)}}=783.0(5.5)~\text{MeV}.
\end{equation}
Here $m_{\eta_s}$ is the mass parameter from the connected diagram and is consistent with the value $m_{\eta_s}=686(4)$ obtained by HPQCD at the physical strange quark mass~\cite{Davies:2009tsa}. This indicates that our sea quark mass parameter is tuned to be almost at the strange quark mass. $m_{\eta_{(1)}}$ is determined from the correlation function that includes the connected diagram and quark annihilation diagram, and is therefore the mass of the physical state $\eta_{(1)}$.

\subsection{Form Factor for $J/\psi\to \gamma^* \eta_{(1)}$}
The transition matrix element $\mathcal{M}$ for the process $J/\psi \to \gamma^* \eta_{(1)}$ can be expressed in terms of one form factor $M(q^2)$, namely,
\begin{eqnarray}\label{eq:matrix}
    \mathcal{M}_{\psi\eta_{(1)}\gamma^*}^\mu &\equiv& \langle \eta_{(1)}(p_{\eta})|j_\mathrm{em}^\mu(0)|\psi(p_{\psi}, \lambda)\rangle \nonumber\\
    &=& M(q^2) \epsilon^{\mu\nu\rho\sigma}p_{\psi,\nu}p_{\eta,\rho}\epsilon_\sigma(p_{\psi},\lambda),
\end{eqnarray}
where $q^2=(p_\psi-p_{\eta})^2$ is the virtuality of the photon, $\epsilon_\sigma(p_{J/\psi},\lambda)$ is the polarization vector of $J/\psi$ and $j_\mathrm{em}^\mu=\bar{c}\gamma^\mu c$ is the electromagnetic current of charm quark (we only consider the initial state radiation and ignore photon emissions from sea quarks and the final state). The matrix element $\mathcal{M}$ is encoded in the following three-point functions 
\begin{equation}\label{eq:three-point}
    C^{\mu i}_{(3)} (\vec{q};t,t')=\sum\limits_{\vec{y}} e^{i\vec{q}\cdot\vec{y}}\langle \mathcal{O}_{\eta}(\vec{p'},t) j_\mathrm{em}^\mu(\vec{y},t')\mathcal{O}_{\psi}^{i,\dagger}(\vec{p},0) \rangle
\end{equation}
with $\vec{q}=\vec{p'}-\vec{p}$, where $\mathcal{O}_{\eta}(\vec{p'},t)$ and $\mathcal{O}_{\psi}^{i}(\vec{p},t)$ are the interpolating field operators for $\eta_{(1)}$ and $J/\psi$ with spatial momenta $\vec{p'}$ and $\vec p$, respectively. For $t\gg t'$, $t'\gg 0$ and in the rest frame of $J/\psi$ ($\vec{p}=0$), the explicit spectral expression of $C^{\mu i}_{(3)}(\vec{q};t,t')$ reads
\begin{eqnarray}\label{eq:spectral}
    C^{\mu i}_{(3)} (\vec{q};t,t')&\approx& \frac{Z_\eta(\vec{q})Z_\psi^{*}}{4V_3 E_\eta(\vec{q})m_{\psi}} e^{-E_{\eta}(\vec{q})(t-t')}e^{-m_{\psi}t'}\nonumber\\
    &\times& \sum\limits_{\lambda}\langle {\eta_{(1)}}(\vec{q})|j_\mathrm{em}^\mu|J/\psi(\vec{0},\lambda)\rangle\epsilon^{*,i}(\vec{0},\lambda),
\end{eqnarray}
where $V_3$ is the spatial volume, $Z_\eta(\vec{q})=\langle \Omega|\mathcal{O}_{\eta_{(1)}}(\vec{q})|{\eta_{(1)}}(\vec{q})\rangle$, and $Z_{\psi}\epsilon^i(\vec{0},\lambda)=\langle \Omega|\mathcal{O}_{\psi}^i(\vec{0})|J/\psi(\vec{0},\lambda)\rangle$. Note that $Z_\eta$ has a $\vec{q}$ dependence due to the smeared operator $\mathcal{O}_{\eta}$~\cite{Bali:2016lva}. The parameters $m_\psi$, $E_\eta(\vec{q})$, $Z_\psi$ and $Z_\eta(\vec{q})$ can be derived from the two-point correlation functions
\begin{eqnarray}\label{eq:two-point}
    C_{(2),\eta}(\vec{q},t)&\approx& \frac{1}{2E_\eta(\vec{q})V_3}|Z_{\eta}(\vec{q})|^2 e^{-E_\eta(\vec{q})t},\nonumber\\
    C_{(2),\psi}^{ii}(t)&\approx&\frac{1}{2m_{\psi}V_3}|Z_{J/\psi}|^2 e^{-m_{\psi}t}.
\end{eqnarray}
Thus we can extract the matrix element $\langle \eta_{(1)}|j_\mathrm{em}^\mu|J/\psi\rangle$ through Eqs.~(\ref{eq:spectral}) and  (\ref{eq:two-point}).

Therefore, the major numerical task is the calculation of $C_{(3)}^{\mu i}(\vec{q};t,t')$. The local EM current $j_\mathrm{em}^\mu(x)=[\bar{c}\gamma^\mu c](x)$ mentioned above (the charm quark field $c$ and $\bar{c}$ are the original field, which are not smeared) is not conserved anymore on the finite lattice and should be renormalized. We determine
the renormalization factor $Z_V^{t}=1.147(1)$ and $Z_V^s=1.191(2)$ for the temporal and spatial components of $j_\mathrm{em}^\mu(x)$,
respectively, by calculating the relevant electromagnetic form factors of $\eta_c$~\cite{Dudek:2006ej,Yang:2012mya}. In practice, only $Z_V^{s}$ is involved and is incorporated implicitly in $j_\mathrm{em}^\mu(x)$. We use the operator $\mathcal{O}_{\gamma_4\gamma_5}=\bar{s}^{(s)}\gamma_4\gamma_5 s^{(s)}$ for $\mathcal{O}_\eta$ and $\mathcal{O}_{\psi}^i$ takes the form $\bar{c}^{(s)}\gamma^i c^{(s)}$ in Eq.~(\ref{eq:three-point}). The three-point function $C_{(3)}^{\mu i}(\vec{q};t,t')$ is calculated in the rest frame of $J/\psi$ ($\vec{p}=0$), such that $\eta_{(1)}$ moves in a spatial momentum $\vec{p}'=\vec{q}$. The right panel of Fig.~\ref{fig:effmass} shows the dispersion relation of $\eta_{(1)}$
\begin{equation}\label{eq:dispersion}
    a_t^2 E_\eta^2(\vec{q})=a_t^2 m_\eta^2 + \frac{1}{\xi^2} \left(\frac{2\pi}{L}\right)^2 |\vec{n}|^2,
\end{equation}
where $\vec{n}$ stands for the momentum mode of $\vec{q}=\frac{2\pi}{La_s} \vec{n}$. It is seen that $E_{\eta}^2(\vec{q})$ exhibits a perfect linear behavior 
in $|\vec{q}|^2$ up to $|\vec{n}|^2=9$ and the fitted slope gives $\xi=4.88(1)$ which deviates from the renormalized anisotropy $\xi\approx 5.0$ by less than 3\%. 

\begin{figure}[t]
    \centering
    \includegraphics[width=0.99\linewidth]{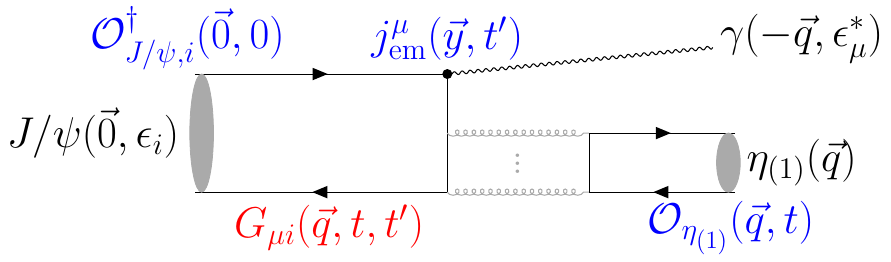}
    \caption{Schematic diagram for the process $J/\psi\to \gamma^* \eta_{(1)}$.}
    \label{fig:quark-diagram}
\end{figure}

After the Wick's contractions, the three-point function $C^{\mu i}_{(3)}$ is expressed in terms of quark propagators, and the schematic quark diagram is illustrated in Fig.~\ref{fig:quark-diagram}. There are two separated quark loops connected by gluons. The strange quark loop on the right-hand side can be calculated in the framework of the distillation method. The left part $G^{\mu i}$ comes from the contraction of $\mathcal{O}_{\psi}^i$ and the current $j^\mu_\mathrm{em}$, namely,
\begin{equation}
G^{\mu i}(\vec{p},\vec{q};t'+\tau,\tau)=\sum\limits_{\vec{y}} e^{i\vec{q}\cdot\vec{y}}j_\mathrm{em}^\mu(\vec{y},t'+\tau)O_{\psi}^{i\dagger}(\vec{p},\tau),
\end{equation}
and is dealt with the distillation method~\cite{Chen:2022isv}. Considering $\mathcal{O}_{\psi}^i(\vec{p},t)=\sum\limits_{\vec{y}}e^{-i\vec{p}\cdot\vec{y}}[\bar{c}^{(s)}\gamma_i c^{(s)}](\vec{y},t)$,  
the explicit expression of $G^{\mu i}$ at the source time slice $\tau=0$ is 
\begin{eqnarray}
    G^{\mu i}(\vec{p},\vec{q};t,0)&=&\sum\limits_{\vec{x}}e^{i\vec{q}\cdot\vec{x}} \mathrm{Tr}\left\{ \gamma_5[S_c V(0)]^\dagger(\vec{x},t)\gamma_5\gamma^\mu\right.\nonumber\\
    &\times& \left. \left[S_c V(0)\right](\vec{x},t)[V^\dagger(0) D(\vec{p})\gamma^i V(0)] \right\},\nonumber\\
\end{eqnarray}

where $S_c=\langle c\bar{c}\rangle_U$ is the all-to-all propagator of charm quark for a given gauge configuration $U$ and $D(\vec{p})$ is a $3L^3\times 3L^3$ diagonal matrix with the diagonal elements being $\delta_{ab} e^{i\vec{p}\cdot\vec{y}}$ ($\vec{y}$ labels the column or row indices and $a,b=1,2,3$ refer to the color indices). The $\gamma_5$-hermiticity $S_c=\gamma_5 S_c^\dagger\gamma_5$ implies $[V^\dagger(0)S_c](\vec{x},t)=\gamma_5 [S_cV(0)]^\dagger(\vec{x},t)\gamma_5$, 
such that only $S_c V(0)$ is required, while $S_c V(0)$ can be obtained by solving the system of linear equations
\begin{equation}
    M[U;m_c][S_c V(0)]=V(0),
\end{equation}
where $M[U;m_c]$ is the fermion matrix in the charm quark action (the linear system solver defined by $M[U;m_c]x=b$ is applied $4N_V$ times for Dirac indices $\alpha=1,2,3,4$ and all the columns of $V(0)$). In order to increase the statistics, the above procedure runs over all the time range, say, $\tau\in[0,T-1]$. Averaging over $\tau\in [0,T-1]$ improves the precision of the calculated $C^{\mu i}_{(3)}$ drastically.

\begin{figure}[t]
    \centering
        \includegraphics[width=0.9\linewidth]{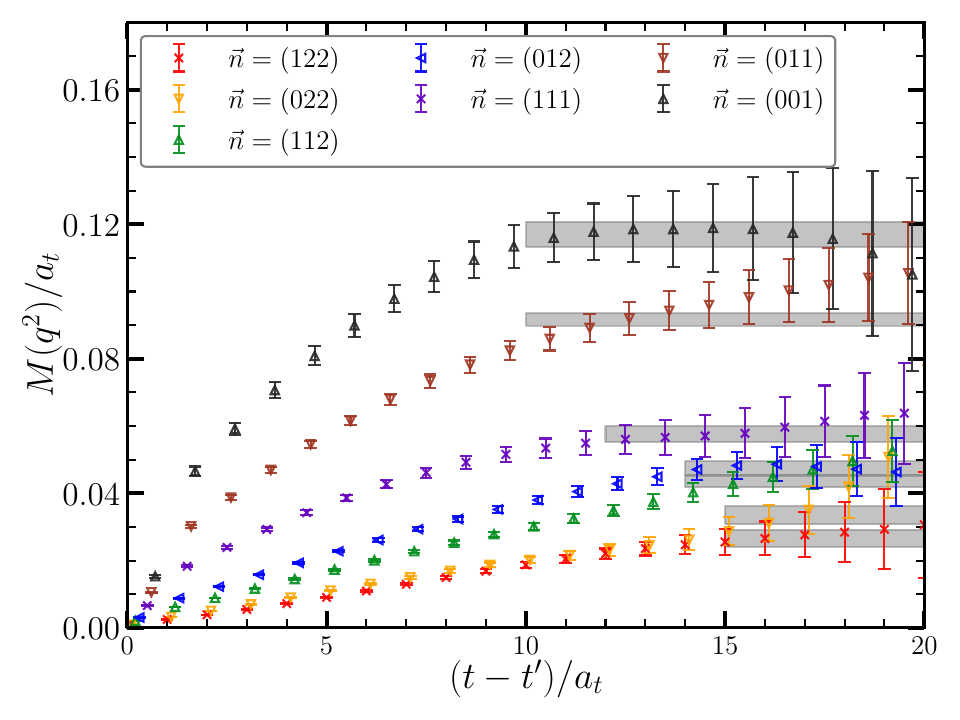}\\
    \caption{Fit of form factor $M(q^2)$ for $J/\psi\to \gamma^* \eta_{(1)}$. The lattice data are plotted as data points and the grey bands show the fit by constants to the plateau regions. The fitted $M(q^2)$ values are given in Table~\ref{tab:fit-values-energy-form-factor}.}
    \label{fig:m-q2}
\end{figure}\

\begin{table*}[t]
    \centering
    \begin{ruledtabular}
        \begin{tabular}{lccccccccc}
         mode $\vec{n}$ of $\vec{q}$ &  $(1, 2, 2)$ & $(0, 2, 2)$ & $(1, 1, 2)$ & $(0, 1, 2)$ 
         & $(1, 1, 1)$ & $(0, 1, 1)$ & $(0, 0, 1)$ \\
         \hline
         $q^2/ \mathrm{GeV}^{2}$        & $-$0.6800(66)& $-$0.1869(73)& 0.8777(91)& 1.459(10)
         & 2.756(14)& 3.499(16)& 4.337(20)\\
         $E_{\eta_{(1)}}^{(\vec{q})} / \mathrm{GeV}$ &  1.803(14) & 1.710(17) & 1.5279(29) & 1.4291(24) 
        & 1.2119(29) & 1.0886(20) & 0.9466(18)\\
         $M(q^2) / \mathrm{GeV}^{-1}$   & 0.00400(38)   & 0.00503(41)   & 0.00654(26)  & 0.00714(30)  
         & 0.00865(36)  & 0.01377(28)  & 0.01757(57)\\
        \end{tabular}
    \end{ruledtabular}
    \caption{Fit values of $\eta_{(1)}$ form factor $M(q^2)$. The momentum modes $\vec{n}$ represent the relation $\vec{q} = \frac{2\pi}{L} \vec{n}$. The two-point function $C_{\Gamma\Gamma}(t)$ and the three-point ratio function $R^{\mu i}(\vec{q};t,t')$ corresponding to the same momentum mode $\vec{n}$ has been averaged for increasing signal of the energy $E_{\eta_{(1)}}^{(\vec{q})}$ and form factor $M(q^2)$, respectively. 
    }
    \label{tab:fit-values-energy-form-factor}
\end{table*}


 It is observed that $J/\psi$ contribution dominates $C^{\mu i}_{(3)} (\vec{q};t,t')$ when $t'>40$. Combining Eqs.~(\ref{eq:matrix},\ref{eq:spectral},\ref{eq:two-point}), we have the following expression
 \begin{eqnarray}
 R^{\mu i}(\vec{q};t,t')&\equiv& \frac{Z_\psi Z_\eta(\vec{q})C^{\mu i}_{(3)}(\vec{q};t,t')}{V_3 C_{(2),\eta}(-\vec{q},t-t')C_{(2),\psi}(t')}\nonumber\\
 &\approx& M(q^2;t-t') \epsilon^{\mu i j} q_j
 \end{eqnarray}
 for the fixed $t'/a_t=40$, from which we obtain $M(q^2,t-t')$ for each $q^2$. 
  Fig.~\ref{fig:m-q2} shows the $t-t'$ dependence of $M(q^2,t-t')$ at several $q^2$ close to $q^2=0$. It is seen that a plateau region appears beyond $t-t'>10$ for each $q^2$, where $M(q^2)$ is obtained through a constant fit. The grey bands illustrate the fitted values and fitting time ranges, along with the jackknife errors. We also test the fit function form $M(q^2,t-t')=M(q^2)+c(q^2) e^{-\delta E (t-t')}$ with the exponential term being introduced to account for the higher state contamination. The fitted values of $M(q^2)$ in this way are consistent with those in the constant fit but have much larger errors. Therefore, we use the results from the constant fit for the values of $M(q^2)$. The derived $M(q^2)$ up to $q^2=4.3~\mathrm{GeV}^2$  data points are list in Table.~\ref{tab:fit-values-energy-form-factor}.
  
 Instead of a polynomial functions form used by Ref.~\cite{Jiang:2022gnd}, we use the single pole model to describe the $q^2$ dependence of $M(q^2)$
 \begin{equation}\label{eq:Dalitz1}
     M(q^2)=\frac{M(0)}{1-q^2/\Lambda^2}\equiv M(0) F_{\psi\eta}(q^2).
 \end{equation}
As indicated by the red band in Fig.~\ref{fig:tff-fit}, the model fits the overall behaviors of $M(q^2)$ very well with the parameters
\begin{eqnarray}\label{eq:result}
    M(0)&=&0.00541(13)~\mathrm{GeV}^{-1},\nonumber\\
    \Lambda&=&2.465(22)~\mathrm{GeV}.
\end{eqnarray}

\section{Discussions}\label{sec:discussion}
\subsection{The partial decay width of $J/\psi\to \gamma \eta_{(1)}$}
The partial decay width $\Gamma(J/\psi\to \gamma \eta_{(1)})$ is dictated by the on-shell form factor $M(q^2=0)$ through the relation
\begin{equation}
    \Gamma(J/\psi\to \gamma\eta_{(1)})=\frac{4\alpha}{27}|M(0)|^2 |\vec{p}_\gamma|^3,
\end{equation}
where the electric charge of charm quark $Q=+2e/3$ has been incorporated, $\alpha\equiv \frac{e^2}{4\pi}=1/134$ is the fine structure constant at the charm quark mass scale, and $|\vec{p}_\gamma|=(m_\psi^2-m_{\eta_{(1)}}^2)/2m_\psi$ is the on-shell momentum of the photon. Using the value of $M(0)$ in Eq.~(\ref{eq:result}), the partial decay width and the corresponding branching fraction are predicted as 
\begin{eqnarray}\label{eq:width}
    \Gamma(J/\psi\to \gamma \eta_{(1)})&=& 0.097(5)~\mathrm{keV}\nonumber\\
    \mathrm{Br}(J/\psi\to \gamma \eta_{(1)})&=& 1.06(5)\times 10^{-3},
\end{eqnarray}
where the experimental value $\Gamma_{J/\psi}=92.6~\mathrm{keV}$ is used. Since an $\eta$ state in the process $J/\psi\to \gamma \eta$ is produced by gluons through its flavor singlet component, the results in Eq.~(\ref{eq:width}) should be compared with the experimental result 
$\mathrm{Br}(J/\psi\to \gamma \eta')=5.25(7)\times 10^{-3}$~\cite{ParticleDataGroup:2022pth} ($\eta'$ is mainly a flavor singlet) and $\mathrm{Br}(J/\psi\to \gamma \eta_{(2)})=4.16(49)\times 10^{-3}$ in the $N_f=2$ case at $m_\pi\approx 350~\mathrm{MeV}$~\cite{Jiang:2022gnd}. Obviously $\mathrm{Br}(J/\psi\to\gamma\eta_{(1)})$ is four or five times smaller that in $N_f=2$ and $N_f=2+1$ cases. 

This large difference can be understood as follows. The decay process $J/\psi\to \gamma\eta_{(N_f)}$ takes place in the procedure that the $c\bar{c}$ pair annihilates into gluons (after a photon radiation), which then convert into 
$\eta_{(N_f)}$. There are two mechanisms for gluons to couple to $\eta_{(N_f)}$. The first is the $\mathrm{U}_A(1)$ anomaly manifested by the anomalous axial vector current relation (in the chiral limit)
\begin{equation}\label{eq:anomaly}
    \partial_\mu j_5^\mu(x)=\sqrt{N_f} \frac{g^2}{32\pi}G_{\mu\nu}^a(x) \tilde{G}^{a,\mu\nu}(x) \equiv \sqrt{N_f} q(x),
\end{equation}
where $j_5^\mu=\frac{1}{\sqrt{N_f}}\sum\limits_{k=1}^{N_f} \bar{q}_k\gamma_5\gamma^\mu q_k$ is the flavor singlet axial vector current for $N_f$ flavor quarks, and $q(x)$ is the topological charge density. The $\mathrm{U}_A(1)$ anomaly induces the anomalous gluon-$\eta$ coupling with the strength observed by the matrix element $\langle 0|q(0)|\eta_{(N_f)}\rangle$. With the matrix element $\langle 0|\partial_\mu j_5^\mu(0)|\eta_{(N_f)}\rangle=f_{\eta_{(N_f)}}m_{\eta_{(N_f)}}^2$, from Eq.~(\ref{eq:anomaly}) one has the relation
\begin{equation}\label{eq:qcoupling}
    \langle 0| q(0) |\eta_{(N_f)}\rangle= \frac{1}{\sqrt{N_f}} f_{\eta_{(N_f)}}m_{\eta_{(N_f)}}^2
\end{equation}
in the chiral limit. According to the Witten and Veneziano mechanism~\cite{Witten:1979vv,Veneziano:1979ec} for the mass of $\eta_{(N_f)}$, $m_{\eta_{(N_f)}}^2=\frac{4N_f}{f_\pi^2} \chi_\mathrm{top}$, where $\chi_\mathrm{top}$ is the topological susceptibility of the SU(3) pure Yang-Mills theory, one has $\langle 0| q(0) |\eta_{(N_f)}\rangle\propto \sqrt{N_f}$ in the chiral limit. 

For massless quarks, the $\mathrm{U}_A(1)$ anomaly dominates the production of $\eta_{(N_f)}$ in the process $J/\psi\to\gamma \eta_{(N_f)}$, 
then one expects the $N_f$ scaling for the partial decay width
\begin{equation}
\Gamma(J/\psi\to\gamma\eta_{(N_f)})\propto |\langle 0|q(0)|\eta_{(N_f)}\rangle|^2\propto N_f,
\end{equation}
since $f_{\eta_{(N_f)}}\approx f_\pi$ is independent of $N_f$ to the lowest order in $1/N_c$. In Ref.~\cite{Jiang:2022gnd}, this scaling relation is used to predict the production rates of $\eta$ and $\eta'$ from the form factor $M(0)$ of the $N_f=2$ case at $m_\pi\approx 350~\mathrm{MeV}$ along with the mixing $\eta-\eta'$ mixing angle $\theta_\mathrm{lin}=-24.5^\circ$~\cite{ParticleDataGroup:2022pth}. The results $\mathrm{Br}(J/\psi\to \gamma\eta)=1.15(14)\times 10^{-3}$ and 
$\mathrm{Br}(J/\psi\to \gamma\eta')=4.49(53)\times 10^{-3}$ are in excellent agreement with the experimental values $1.11(3)\times 10^{-3}$ and $5.25(7)\times 10^{-3}$, respectively. For the case of this study of $N_f=1$ strange quarks, the scaling relation implies $\mathrm{Br}(J/\psi\to \gamma\eta_{(N_f=3)}) = 3.19(15)\times 10^{-3}$ which is closer to the experiment value of $J/\psi\to \gamma \eta'$. This is in the right trend but still has a large discrepancy that can be attributed to the quark mass dependence. The quark mass dependence appears in three places.
First, there is an additional term $2m_s j_5\equiv 2m_s (i\bar{s}\gamma_5 s)$ in the right-hand side of Eq.~(\ref{eq:anomaly}) for $N_f=1$ strange quarks, which gives a correction to Eq.~(\ref{eq:qcoupling}) as 
\begin{equation}
    \langle 0| q(0) |\eta_{(1)}\rangle= f_{\eta_{(N_f)}}m_{\eta_{(N_f)}}^2-2m_s \langle 0|j_5(0)|\eta_{(1)}\rangle
\end{equation}
with $m_s \langle 0|j_5(0)|\eta_{(1)}\rangle>0$~\cite{Novikov:1979uy,Feldmann:1999uf,Beneke:2002jn,Cheng:2008ss,Singh:2013oya,Qin:2017qes,Ding:2018xwy,Bali:2021qem}.
Secondly, the right-hand side of Eq.~(\ref{eq:qcoupling}) has quark mass dependence itself through $f_{\eta_{(N_f)}}$ and $m_{\eta_{(N_f)}}$. Since the strange quark is not so light as $u,d$ quarks, this kind of the quark mass dependence may not be negligible. The third place is the $\eta_{(N_f)}$
production procedure that the gluons couple perturbatively to $q\bar{q}$ pairs which then couple to $\eta_{(N_f)}$. Because of the vector-like $q\bar{q}$-gluon coupling in QCD, the produced quark and antiquark have opposite chiralities. So in the chiral limit, the production of $\eta_{(N_f)}$ is prohibited to all orders of the perturbative QCD due to the conservation of angular momentum (the $q\bar{q}$ pair has chirality two and has a total spin $S=0$ and the orbital angular momentum $L=0$). For massive quarks, the amplitude of this process is proportional to the quark mass~\cite{Chanowitz:2005du,Chao:2007sk}, which measures the mixing of quarks with different chiralities. Thus the production of $s\bar{s}$ is drastically enhanced over that of $(u\bar{u}+d\bar{d})/\sqrt{2}$. If the amplitude of this mechanism for the $\eta_{(N_f)}$ production has an opposite 
sign to that of the $\mathrm{U}_A(1)$ anomaly, then the seemingly smaller value in Eq.~(\ref{eq:width}) can be understood. This possibility actually exists, as is manifested in a previous lattice QCD study on the semileptonic decay from $D_s$ to $\eta'$~\cite{Bali:2014pva}, where it is observed that, the contribution of the direct coupling of $s\bar{s}$ to $\eta'$ has an opposite sign to that through the disconnected diagram.  
\begin{figure}[t]
    \centering
    \includegraphics[width=0.95\linewidth]{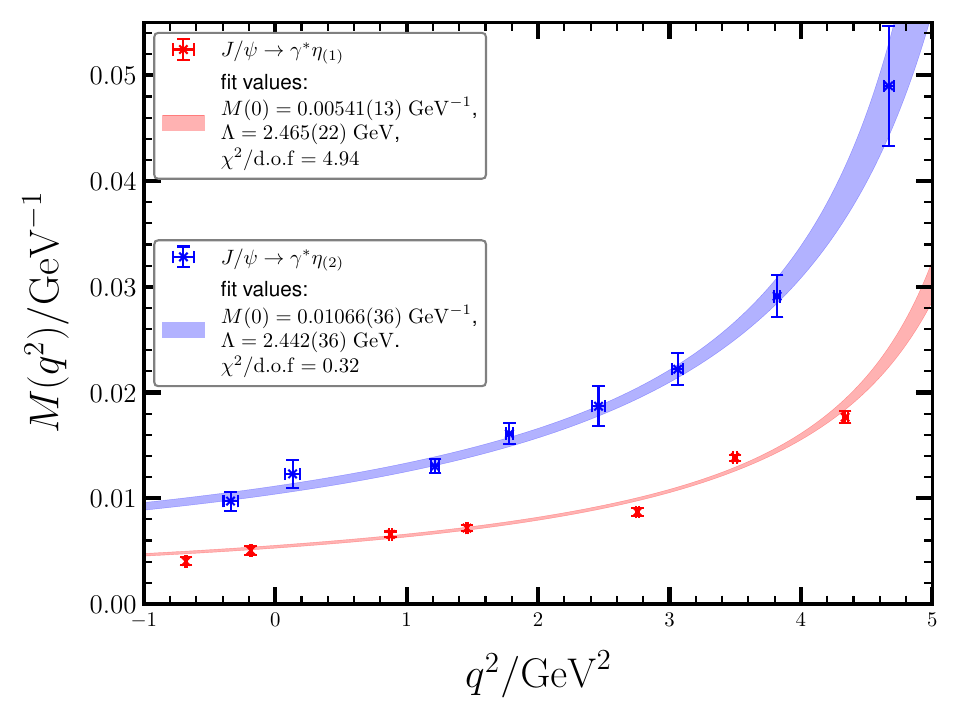}\\
    \caption{The form factor $M(q^2)$ for $J/\psi\to \gamma^* \eta_{(1,2)}$. The data points are the lattice QCD results, and the shaded 
    bands illustrate the fit model $M(q^2)=\frac{M(0)}{1-q^2/\Lambda^2}$ with the best fit parameters $M(0)=0.01066(36)~\text{GeV}^{-1}$ for $N_f=1$ and $\Lambda=2.442(36)~\text{GeV}$ for $N_f=2$. The $M(q^2)$ data of $J/\psi\to\gamma^*\eta_{(2)}$ are the same as those in Table II of Ref.~\cite{Jiang:2022gnd} and the fit is performed using the jackknife method on the original data sample.  
    }
    \label{fig:tff-fit}
\end{figure}
\subsection{The Dalitz decay form factors $J/\psi\to P l^+l^-$}
The form factor $M(q^2)$ in Eq.~(\ref{eq:Dalitz1}) is actually the TFF for the Dalitz decay $J/\psi\to\eta_{(1)} l^+l^-$ when $q^2>4m_l^2$, which is seen to be well described by the single pole model with $\Lambda=2.465(22)~\mathrm{GeV}$. 
In Ref.~\cite{Jiang:2022gnd}, the Dalitz TFF $M(q^2)$ is also obtained in the $N_f=2$ lattice QCD at $m_\pi\approx 350~\mathrm{MeV}$, and the value $M(0)$ is interpolated using a polynomial function form. We refit the $q^2$-dependence of $M(q^2)$ using the same single pole model, as shown in Fig.~\ref{fig:tff-fit}. It is observed that the single pole model describes the data better than the polynomial model in the whole $q^2$ range,
and the pole parameter $\Lambda=2.442(36)~\mathrm{GeV}$ agrees well with the value for $N_f=1$. This signals the single pole model may be universal
for the Dalitz decays from $J/\psi$ to light pseudoscalar mesons $P$ and the pole parameter $\Lambda$ is insensitive to the number of light flavors $N_f$ and the mass of $P$. 

In experiments, the TFF $F_{\psi P}$ can be extracted from the ratio
\begin{equation}\label{eq:ratio}
    \frac{d\Gamma(\psi\to P l^+ l^-)/dq^2}{\Gamma(\psi\to P \gamma)} = A(q^2) |F_{\psi P}(q^2)|^2,
\end{equation}
where $A(q^2)$ is a known kinematic factor~\cite{Landsberg:1985gaz,Fu:2011yy,Gu:2019qwo} 
\begin{eqnarray}
A(q^2)&=&\frac{\alpha}{3\pi}\frac{1}{q^2} \left(1-\frac{4m_l^2}{q^2}\right)^{1/2}\left( 1+\frac{2m_l^2}{q^2}\right)\nonumber\\
              &\times& \left[ \left(1+\frac{q^2}{m_\psi^2-m_P^2}\right)^2-\frac{4m_\psi^2 q^2}{(m_\psi^2-m_P^2)^2} \right]^{3/2}
\end{eqnarray}
derived from the QED calculation. 
BESIII has measured many Dalitz decay processes of $J/\psi\to P e^+e^-$ with $P=\eta$~\cite{BESIII:2014dax,BESIII:2018qzg}, $\eta'$~\cite{BESIII:2014dax,BESIII:2018iig,BESIII:2018aao}, $\eta(1405)$~\cite{BESIII:2023jyg}, and $(X(1835),X(2120),X(2370))$~~\cite{BESIII:2021xoh}. For some of these processes, the TFF are obtained and fitted through the single pole model (along with resonance terms if experimental data are precise enough~\cite{BESIII:2018qzg}) in Eq.~(\ref{eq:pole}) and the fitted values of $\Lambda$ are collected in Table~\ref{tab:Lambda}, where the values of $\Lambda$ derived from lattice QCD are also presented in the last two rows for comparison. Although the values of $\Lambda$ for the $J/\psi\to \eta,\eta'$ Dalitz decays are compatible with the lattice values, the values of $\Lambda$ for $J/\psi\to \eta(1405), X(1835)$ are substantially smaller. So it is possible that $\Lambda$ depends on the mass of the final state pseudoscalar meson.   
\begin{table}[t]
    \centering
    \caption{The values of the pole parameter $\Lambda$ of the TFF for different Dalitz decays $J/\psi\to P e^+ e^-$. The $N_f=1,2$ lattice QCD results of $\Lambda$ are also shown in the bottom two rows for comparison. 
    \label{tab:Lambda}}
    \begin{ruledtabular}
    \begin{tabular}{llc}
        $V\to e^+ e^- P$                & $\Lambda~(\text{GeV})$    & Ref. \\\hline
        $J/\psi\to e^+e^- \eta$         & $2.56\pm 0.04\pm 0.03$    & {}~\cite{BESIII:2018qzg}\\
        $J/\psi\to e^+e^- \eta'$        & $3.1~\pm 1.0$             & {}~\cite{BESIII:2014dax}\\
        $J/\psi\to e^+e^- \eta(1405) $  & $1.96\pm 0.24\pm 0.06$    & {}~\cite{BESIII:2023jyg}\\
        $J/\psi\to e^+e^- X(1835) $     & $1.75\pm 0.29\pm 0.05$    & {}~\cite{BESIII:2021xoh}\\
        \hline
        $J/\psi\to \gamma^* \eta_{(2)}(718)$ & $2.44\pm 0.04$            & {}~\cite{Jiang:2022gnd}\\
        $J/\psi\to \gamma^* \eta_{(1)}(783)$ & $2.47\pm 0.02$            & this work
    \end{tabular}
    \end{ruledtabular}
\end{table}

\begin{figure}[t]
    \centering
    \includegraphics[width=1.0\linewidth]{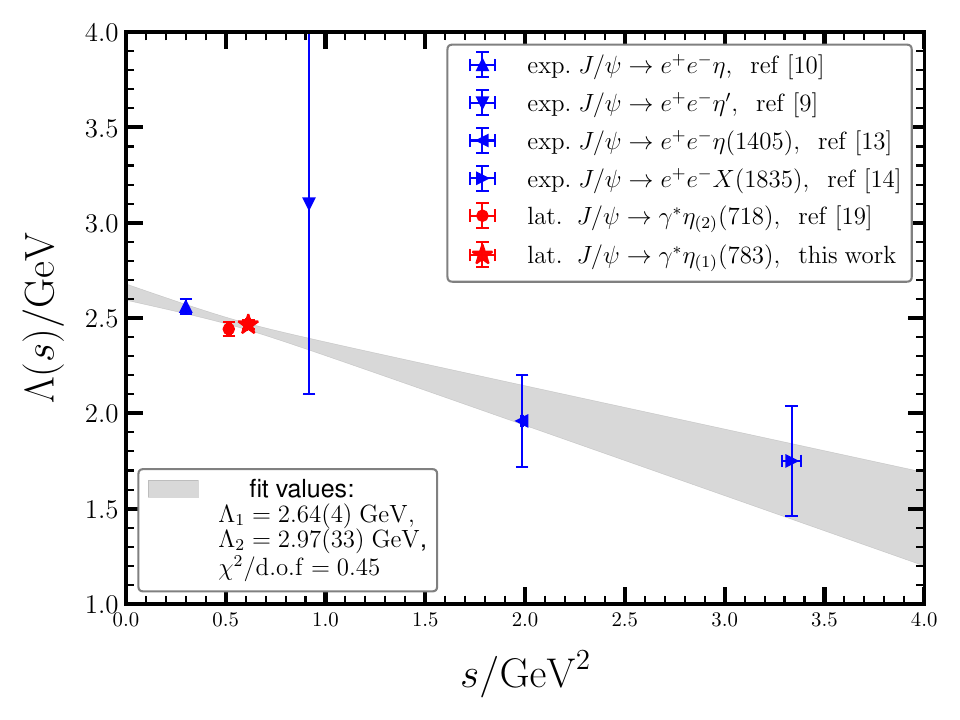}
    \caption{The $s$ dependence of the pole parameter $\Lambda$. The data points indicate the experimental and lattice result of $\Lambda$ at different values of $s=m_P^2$ (listed in Table~\ref{tab:Lambda}), where $m_P$ is the mass of the psuedoscalar meson in the process $J/\psi\to P\gamma^*$. The grey band shows the model $\Lambda(s) = \Lambda_1(1 - \frac{s}{\Lambda_2^2})$ with the fitted parameters $\Lambda_1 = 2.64(4)~\mathrm{GeV}$ and $\Lambda_2 = 2.97(33)~\mathrm{GeV}$. The $\chi^2$ per degree of freedom is $\chi^2/\mathrm{d.o.f}=0.45$.}
    \label{fit-form-factor}
\end{figure}

In principle, the production of each light pseudoscalar $P$ in the $J/\psi$ radiative decay or the Dalitz decay undergoes the same procedure that the $c\bar{c}$ pair emits a photon of the virtuality $q^2$ and then annihilates into gluons, whose invariant mass squared is labelled as $s$. Since the single pole model describes $M(q^2)$ very well while the $q^2$ and $s$ in the $J/\psi-\gamma^*(q^2)-(gg\cdots)^*(s)$ vertex are correlated, one expects the the $s$-dependence of $\Lambda$. We assume a linear function form for $\Lambda(s)$
\begin{equation}\label{eq:lambdas}
    \Lambda(s)=\Lambda_1\left(1-\frac{s}{\Lambda_2^2}\right).
\end{equation}
Then using the values of $\Lambda$ in Table~\ref{tab:Lambda} that are measured from experiments and lattice QCD studies at different $s=m_P^2$, the parameters $\Lambda_1$ and $\Lambda_2$ can be fitted through the above equation. Finally, we get 
\begin{equation}\label{eq:2pole-para}
    \Lambda_1=2.64(4)~\mathrm{GeV}, ~~~\Lambda_2=2.97(33)~\mathrm{GeV}
\end{equation}
with $\chi^2$ per degree of freedom $\chi^2/\mathrm{d.o.f}=0.45$. The values of $\Lambda_{1,2}$ in Eq.~(\ref{eq:2pole-para}) can give inputs for theoretical and experimental studies. Taking the process $J/\psi\to \eta' e^+e^-$ for instance, the experimental value of $\Lambda$ has huge uncertainties, but the model in Eq.~(\ref{eq:lambdas}) with the parameters in Eq.~(\ref{eq:2pole-para}) gives a more precise prediction
\begin{equation}
    \Lambda(s=m_{\eta'}^2)=2.36(3)~\mathrm{GeV}.
\end{equation}
Then according to Eq.~(\ref{eq:ratio}) and using the experimental result of $\mathrm{Br}(J/\psi\to\gamma \eta')=5.25(7)\times 10^{-3}$, the branching fraction of 
$J/\psi\to \eta' e^+e^-$ is estimated to be $6.05(3)(8)\times 10^{-5}$, which is compatible with the BESIII result $6.59(7)(17)\times 10^{-5}$~\cite{BESIII:2018aao}. When the $\rho$ resonance contribution is included, as did by BESIII for $J/\psi\to \eta e^+e^-$ in Ref.~\cite{BESIII:2018qzg}, the $|F_{\psi\eta'}(q^2)|^2$ reads
\begin{eqnarray}
|F_{\psi\eta'}(q^2)|^2&=&|A_\rho|^2\left(\frac{m_\rho^4}{(q^2-m_\rho^2)^2+m_\rho^2\Gamma_\rho^2}\right)\nonumber\\
&+& |A_\Lambda|^2 \left(\frac{1}{1-q^2/\Lambda^2}\right)^2,
\end{eqnarray}
where $A_\rho$ is the coupling constant of the $\rho$ meson and $A_\Lambda$ is the coupling constant of the non-resonant contribution. For $J/\psi\to\eta e^+e^-$, BESIII determines $A_\rho=0.23(4)$ and $A_\Lambda=1.05(3)$~\cite{BESIII:2018qzg}, which give $|F_{\psi\eta}(q^2\approx 0)|^2=1.11\pm 0.07\pm 0.07$. If we take the same value for $A_\rho=0.23(4)$ and assume $A_\Lambda=1$ for the case of $\eta'$ (The $|F_{\psi\eta'}(q^2)|^2$ at $q^2\approx 0$ in Ref.~\cite{BESIII:2014dax} is consist with one within errors), then using the PDG values of $m_\rho$ and $\Gamma_\rho$~\cite{Colquhoun:2023zbc} we get 
\begin{equation}
    \mathrm{Br}(J/\psi\to \eta' e^+e^-)=6.58_{-17}^{+21}(2)(9)\times 10^{-5},
\end{equation}
where the first error is is due to the uncertainty of $A_\rho$, the second is from that of $\Lambda$, and the third is from that of the experimental value of $\mathrm{Br}(J/\psi\to\gamma\eta')$. This value agrees with the experimental value better. 

\section{Summary}\label{sec:summary}
We generate a large gauge ensemble with $N_f=1$ dynamical strange quarks on an anisotropic lattice with the anisotropy $a_s/a_t\approx 5.0$. The pseudoscalar mass is measured to be $m_{\eta_s}=693.1(3)~\mathrm{MeV}$ without considering the quark annihilation effect, and $m_{\eta_{(1)}}=783.0(5.5)~\mathrm{MeV}$ with the inclusion the quark annihilation diagrams. We calculate the EM form factor $M(q^2)$ for the decay process $J/\psi\to \gamma^*(q^2) \eta_{(1)}$ with $q^2$ being the virtuality of the photon. By interpolating $M(q^2)$ to the value at $q^2=0$ through the VMD inspired single pole model in Eq.~(\ref{eq:Dalitz1}), the decay width and the branching fraction of $J/\psi\to \gamma \eta_{(1)}$ is predicted to be $\Gamma(J/\psi\to \gamma\eta_{(1)})=0.097(5)~\mathrm{keV}$
and $\mathrm{Br}(J/\psi\to \gamma\eta_{(1)})=1.06(5)\times 10^{-3}$, respectively, which are much smaller than those in the $N_f=2$ case and those in the physical $N_f=2+1$ case. 
The difference among the partial widths $\Gamma(J/\psi\to \gamma \eta_{(N_f)})$ at different $N_f$ can be attributed in part to the $\mathbf{U}_A(1)$ anomaly that induces a $N_f$ scaling. 

It is interesting to see that $M(q^2)$'s in $N_f=1,2$ are both well described by the single pole model $M(q^2)=M(0)/(1-q^2/\Lambda^2)$. Combined together with the known experimental results of the Dalitz decays $J/\psi\to Pe^+e^-$ with $P$ being light pseudoscalar mesons, the $m_P$ dependence of the pole parameter
$\Lambda$ is observed and can be expressed approximately as $\Lambda(m_P^2)=\Lambda_1(1-m_P^2/\Lambda_2^2)$ with $\Lambda_1=2.64(4)~\mathrm{GeV}$ and $\Lambda_2=2.97(33)~\mathrm{GeV}$. This result provide meaningful inputs for future theoretical and experimental studied on Dalitz decays $J/\psi\to Pe^+e^-$. As a direct application, this $m_P$ dependence expects a pole parameter $\Lambda(s=m_{\eta'}^2)=2.36(3)~\mathrm{GeV}$, which is more precise than the value $3.1(1.0)~\mathrm{GeV}$ measured by BESIII~\cite{BESIII:2014dax} and whose prediction on $\mathrm{Br}(J/\psi\to \eta'e^+e^-)$ agrees better with the experimental value~\cite{BESIII:2018aao}.
\begin{acknowledgments}
This work is supported by the National Natural Science Foundation of China (NNSFC) under Grants No. 11935017, No. 12293060, No. 12293065, No. 12293061, No. 12293062, No. 12293063, No. 12075253, No. 12192264, No. 12175063, No. 12205311, No. 12070131001 (CRC 110 by DFG and NNSFC)), and the National Key Research and Development Program of China (No. 2020YFA0406400) and the Strategic Priority Research Program of Chinese Academy of Sciences (No. XDB34030302). The Chroma software system~\cite{Edwards:2004sx} and QUDA library~\cite{Clark:2009wm,Babich:2011np} are acknowledged. The computations were performed on the HPC clusters at Institute of High Energy Physics (Beijing) and China Spallation Neutron Source (Dongguan), and the ORISE computing environment.
\end{acknowledgments}

\bibliography{ref}
\end{document}